\documentclass[a4paper]{jpconf}
\usepackage{graphicx}
\begin{document}
\title{Neutrinos : from the r-process to the diffuse supernova neutrino background}

\author{Maria Cristina Volpe}

\address{Astro-Particule et Cosmologie (APC), CNRS UMR 7164, Universit\'e Denis Diderot,\\ 10, rue Alice Domon et L\'eonie Duquet, 75205 Paris Cedex 13, France}

\ead{volpe@apc.univ-paris7.fr}

\begin{abstract}
Neutrinos from dense environments are connected to the longstanding open questions of how massive stars explode and what are the sites where $r$-process elements are made.
Flavor evolution and neutrino properties can influence nucleosynthetic abundances.  
GW170817 has given indirect evidence for r-process elements in binary neutron star mergers. 
We discuss the impact of non-standard interactions in such sites.
Nearby compact objects, strong gravitational fields are present. We discuss their influence upon neutrino decoherence in a wave packet treatment of neutrino propagation. 
We conclude by mentioning the upcoming measurement of the diffuse supernova neutrino background.

\end{abstract}

\section{Introduction}
\subsection{Core-collapse supernovae and binary neutron star mergers}
There are powerful sources of low energy neutrinos in the Universe. Iron core-collapse supernovae are massive stars (M $>$ 8 M$_{\rm sun}$) that produce $10^{58}$ $\nu$  in 10 seconds. Neutrinos of all flavors takes away $99 \%$ of the gravitational binding energy (a few $10^{53}$ ergs).  With comparable luminosities, binary neutron star merger  (BNS) remnants produce $\nu_e$, $\nu_{\mu}$ and $\nu_{\tau}$ as well as antineutrinos of all flavors in the tens of MeV energy range.

Evidence for supernova neutrinos was brought by Sanduleak 69$^{\circ}$202 that exploded at 50 kpc (in the Large Magellanic Cloud), giving SN1987A. This is a unique event to date where neutrinos from the inner core of an exploding massive star were observed. A total number of 24 neutrino events were recorded in Kamiokande \cite{Hirata:1987hu}, IMB \cite{Bionta:1987qt} and Baksan \cite{Alekseev:1988gp} (LVD events \cite{Aglietta:1987it} remain debated), with energy spectra and fluences in good agreement with expectations \cite{Vissani:2014doa}.
This exceptional observation refuted the {\it prompt explosion} model and favored the {\it delayed neutrino heating mechanism} by Bethe and Wilson \cite{Bethe:1985sox} that includes an accretion phase \cite{Loredo:2001rx}. The latter is currently considered the mechanism under which the majority of core-collapse supernovae explode. SN1987A events brought a wealth of information on astrophysics, on neutrino properties (e.g. neutrino speed and magnetic moment), non-standard particles like axions, and interactions. Moreover the observations of asymmetric ejecta and strong mixing gave momentum to multidimensional simulations that nowadays (over)explode in 2D and seem to be close to successfull explosions in 3D \cite{Janka:2017vcp,Radice:2017kmj,Bruenn:2018wpz}.  

GW170817 \cite{LIGOScientific:2017vwq} is a unique event where gravitational waves from binary neutron star mergers were measured, in coincidence with a short gamma ray burst and a kilonova. The electromagnetic signal, in comparison with models, shows presence of lanthanides-free ejecta (blue component) and ejecta with lanthanides (red component). This constitutes indirect evidence for the production of $r$-process elements in binary neutron star mergers which are candidate sites, as core-collapse supernovae, for $r$-process nucleosynthesis\footnote{The rapid neutron capture process, responsible for the production of about half of the elements heavier than iron in our solar system, occurs on timescales of seconds and requires a neutron rich environment.}. 
Neutrino properties and flavor evolution impact $r$-process nucleosynthetic abundances. 

An established flavor mechanism is the Mikheev-Smirnov-Wolfenstein (MSW) effect \cite{Wolfenstein:1977ue,Mikheev:1986gs}, due to neutrino coupling with matter, that is responsible for the suppression of solar $^{8}$B neutrinos.  Important progress has been done in the last two decades in the understanding of how neutrinos change flavor in dense astrophysical environments. A wealth of theoretical investigation has clearly shown that neutrinos change their flavor in astrophysical and cosmological environments due to standard and non-standard interactions, to neutrino self-interactions, shock waves and turbulence (in core-collapse supernovae). Flavor mechanisms are intertwined with neutrino properties. The inclusion of neutrino mixings and of the full collision term for cosmological neutrinos as well as of plasma QED corrections has given a very precise value of the effective number of degrees of freedom $N_{ \rm eff} = 3.0440$ \cite{Froustey:2020mcq}.
Nearby compact objects (black holes or neutron stars), neutrinos evolve in presence of strong gravitational fields. These influence neutrino evolution through the redshift, the bending of the trajectories (see e.g. \cite{Yang:2017asl,Caballero:2011dw}) and by modifying interference and flavor phenomena due to decoherence in curved spacetime \cite{Chatelain:2019nkf}. 

These theoretical studies are essential to unravel the sites and conditions for $r$-process nucleosynthesis, to determine the core-collapse supernova mechanism and for future observations of supernova neutrinos and of the diffuse supernova neutrino background.

 \section{Non-standard interactions and flavor evolution in dense media} 
Non-standard interactions are present in theories beyond the standard model. Constraints on non standard neutrino-matter interactions (NSI) are given by solar, oscillations and scattering experiments (see e.g. \cite{Davidson:2003ha,Farzan:2017xzy}).
Further limits have been obtained by neutrino nucleus coherent scattering \cite{COHERENT:2017ipa}, using both time and energy information for example in \cite{Giunti:2019xpr}. 
 Non standard neutrino-neutrino interactions constraints are loose \cite{Bilenky:1999dn}.
Apart from giving different values of the mixing parameters, if NSI exist they strongly influence neutrino flavor evolution in dense environments.

\begin{figure}\label{fig:nsi}
\begin{center}
\includegraphics[width=.3\textwidth]{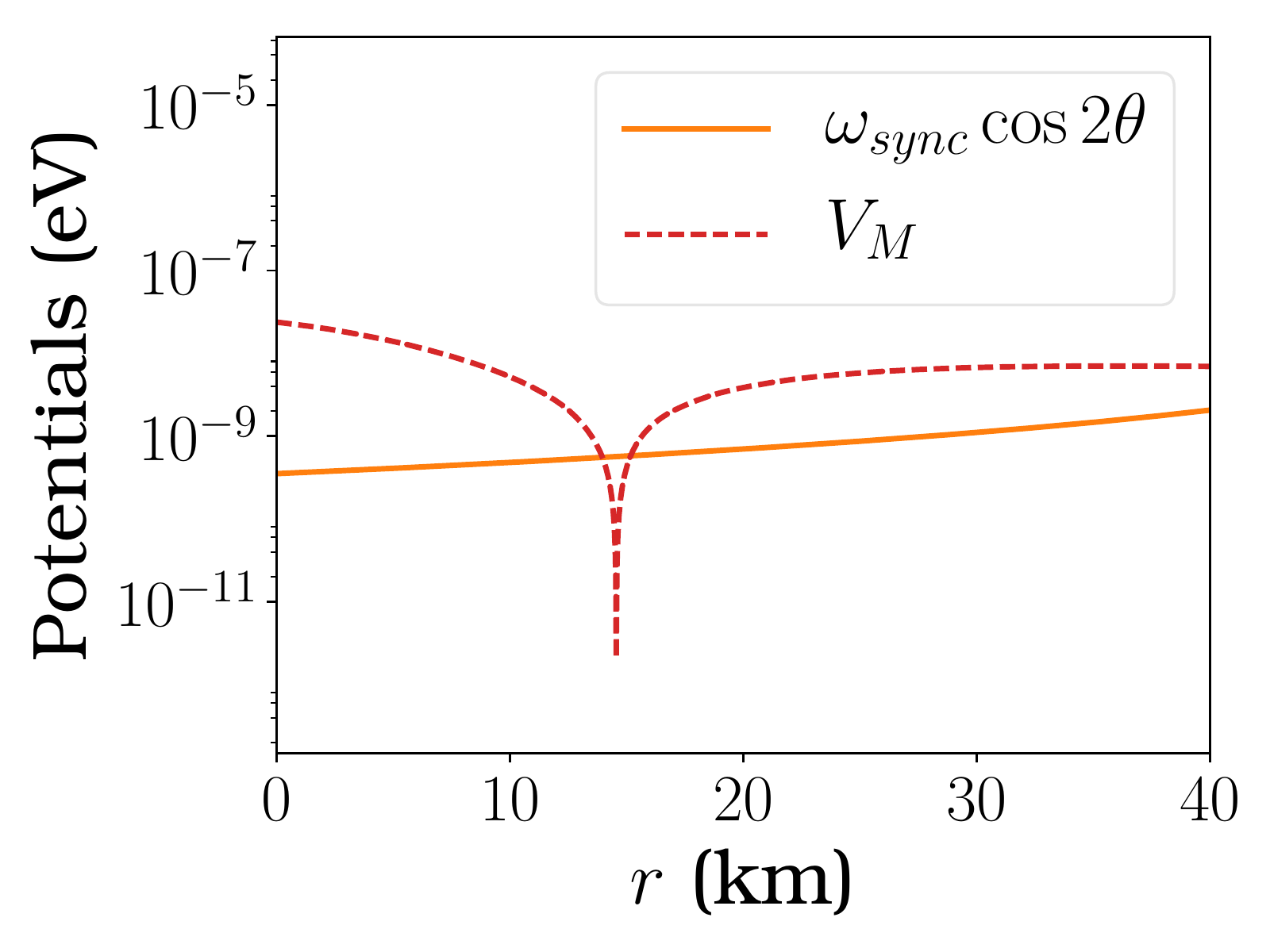}
\includegraphics[width=.3\textwidth]{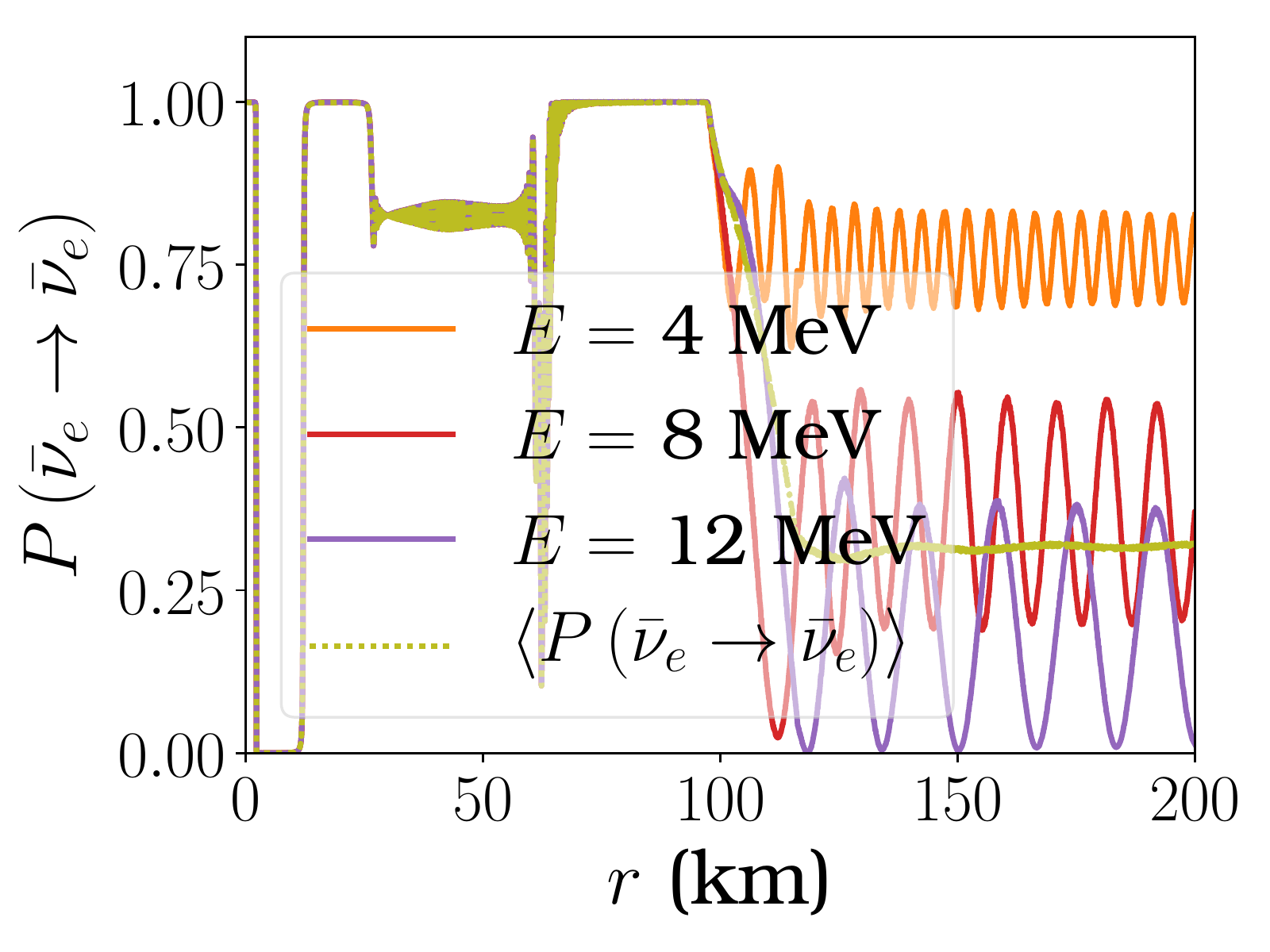}
\includegraphics[scale =.3]{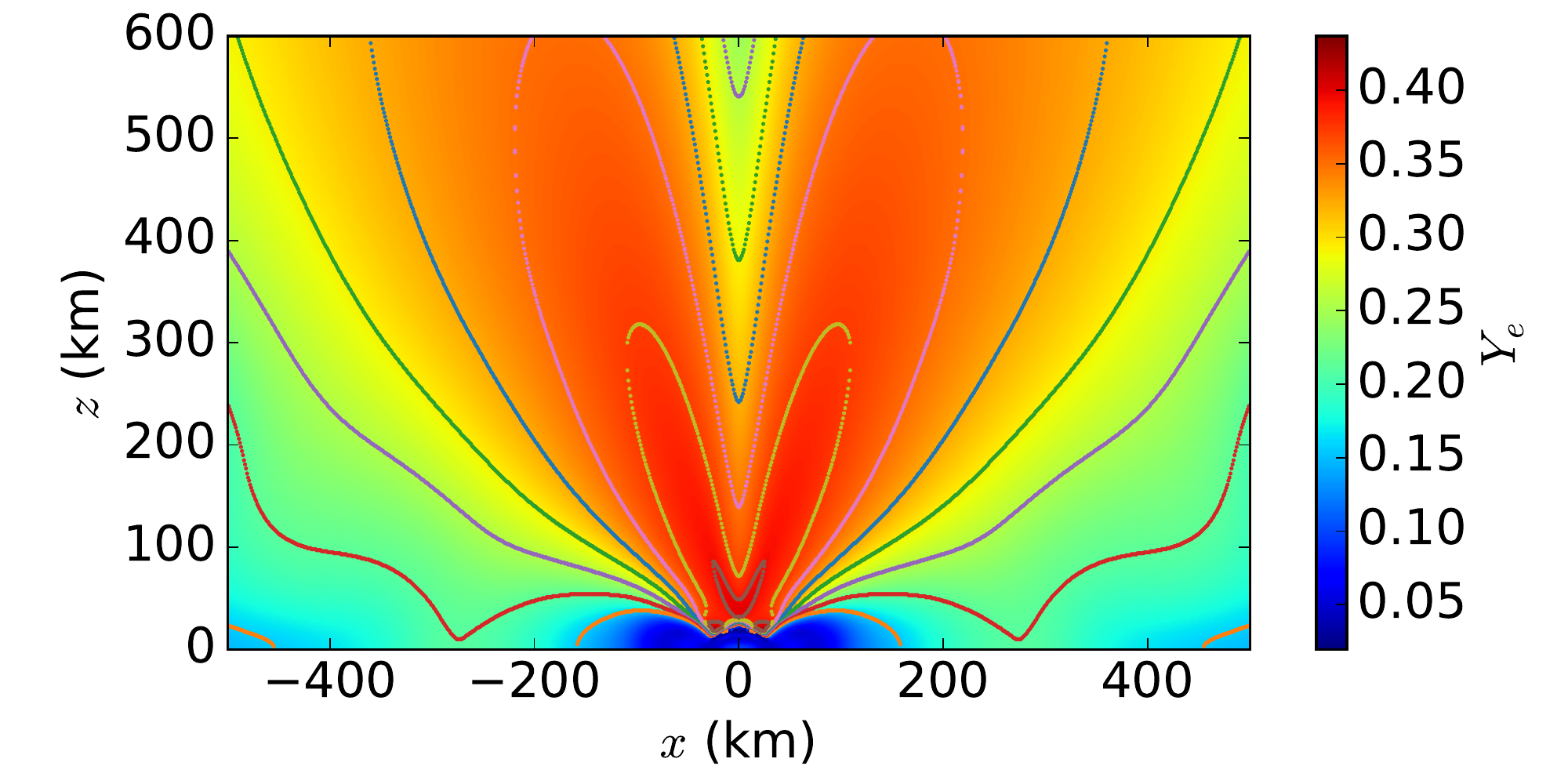}
\caption{Neutrino flavor evolution in an a BNS merger in presence of non-standard neutrino matter interactions. Left : Both the (standard and non-standard) matter potential $V_M$ and the effective spins precession frequency $\omega_{sync}$, inversely proportional to the self-interaction coupling, are shown. The crossing of the two curves corresponds to the synchronized I-resonance condition. The results are for $\epsilon_0 = 10^{-4}, \delta \epsilon^n  = -0.88$. Middle : Electron antineutrino survival probability for $\epsilon_0 = 10^{-4}$, $\delta \epsilon^n  = - 0.9$. Neutrinos undergo four I-resonances, the first is a "normal" one, whereas the other three are synchronized. At about 100 km a MNR resonance takes place. Right : The curves show locations where the I-resonance condition is met. Note that the $Y_e$ values here are from the BNS simulations, without flavor evolution. Adapted from \cite{Chatelain:2017yxx}.}
\end{center}
\end{figure}

In the core-collapse supernova context, NSI have been studied for example in 
\cite{Esteban-Pretel:2007zkv,Stapleford:2016jgz}. Ref.\cite{Fogli:2002xj} studied the influence on the MSW H- and L-resonances. Ref. \cite{Esteban-Pretel:2007zkv} investigated the interplay of non-standard interactions and neutrino self-interactions\footnote{In the so called {\it bulb model}, assuming spherical symmetry.}. A new MSW-like phenomenon, the I-resonance\footnote{I stands for inner} takes place, in the inner deleptonized stellar layers, due to a cancellation of the standard and non-standard terms. Ref.\cite{Stapleford:2016jgz} has performed an extensive investigation of the NSI in presence of self-interactions covering a wide range of NSI parameters, showing the appearance of an MSW-like resonance,  the matter neutrino resonance (MNR), previously seen in BNS\footnote{The MNR occurs because of a cancellation between the matter and the neutrino self-interaction terms.} \cite{Malkus:2012ts}.

Ref.\cite{Chatelain:2017yxx} has performed the first study of NSI effects (in two flavors)  in BNS based on the multidimensional simulations of BNS remnants (from Ref.\cite{Perego:2014fma}).
Contrarily to previous findings, this study uncovered that the I-resonance can be triggered by self-interactions. By using the formalism of effective spins, the effective neutrino and antineutrino spins are shown to go through the resonance simultaneously and independently of their energy. The spin precession frequency $\omega_{sync}$, inversely proportional to the neutrino self-interaction coupling, gets small for sizeable self-interactions and meets the smallness of the  (standard and non-standard) matter term (Figure 1). 

The investigation of a large number of trajectories (with NSI coupling values below current bounds) has uncovered a complex interplay of flavor mechanisms, producing neutrino spectral modifications. Interestingly significant flavor modification is found even for very small values of the NSI couplings. An example is given in Figure 1, which also shows that the I-resonance condition is fullfilled both in the BNS funnel and in the polar regions. While a self-consistent calculation of a neutrino driven wind following the matter composition and flavor evolution is still missing, such studies show that NSI can clear impact r-process nucleosynthetic abundances in kilonovae \cite{Chatelain:2017yxx} and in core-collapse supernovae \cite{Stapleford:2016jgz}. 

\section{Decoherence by wave packet separation due to strong gravitational fields}
In most studies, neutrinos are treated as plane waves. Neutrinos being localized particles a wave packets (WPs) description would be necessary, although this increases the complexity of the theoretical description of neutrinos in dense environments even further. In such a framework, if the WPs associated with each mass eigenstate overlap at a production point, they separate during propagation and their decoherence can suppress flavor modification. 

WPs decoherence can be quantified by the decoherence length $L_{coh}$, distance at which the separation between the two WPs centroids (in two neutrino flavors) satisfies $ \Delta x = \sigma_x$, with $\sigma_x$ the WP width that depends on the production process (see e.g. \cite{Kersten:2015kio}). Neutrino decoherence by WPs separation has been studied in flat spacetime (see e.g. \cite{Kersten:2015kio,Akhmedov:2017mcc,Giunti:2003ax}). In a density matrix description of neutrino propagatioin, for Gaussian WPs, one can show that the off-diagonal elements $\rho_{jk}$ of the neutrino density matrix are exponentially damped with  $L^{jk}_{coh} = 4\sqrt{2} E^2 \sigma_x / \vert \Delta m^2_{jk} \vert $, $E$ being the neutrino energy and $ \Delta m^2_{jk}$ the squared mass differences between the $j, k$ mass eigenstates. 

Nearby compact objects, the presence of strong gravitational fields can modify WPs decoherence. This was studied for the first time in
Ref. \cite{Chatelain:2019nkf} that has extended the density matrix formalism to curved spacetime (Figure 2). The case of a static gravitational field with spherical symmetry, described by the Schwarzschild metric, was considered. It was shown that for Gaussian WPs, the off-diagonal elements of the density matrix have a similar factorisation as in the flat spacetime case, with an exponential suppression whose damping factor also has analogous expression as in flat spacetime. However, it depends now on the (unphysical) coherence coordinate distance $r^{PD}_{coh}$ and the neutrino energy for an observer at infinity. 

Using kinematical arguments, one can relate the $r^{PD}_{coh}$ to a coherence proper time $\tau_{coh}$ at which the difference in the coherence times at a "detection" point D satisfies $\tau_{D}^{jk} = \sigma_t \sqrt{B(r_D)}$ with $\sigma_x \approx \sigma_t$, $B(r) = 1 - r_s/r$, $r_s = 2M$ being the Schwarzschild radius, $M$ the compact object mass. WPs decoherence, with respect to flat spacetime, i.e. $\eta = (\tau_{coh} - L_{coh})/L_{coh}$ (in percentage) is shown in Figure 2. One can see a modification of several tens of a percent for typical neutron star masses, showing that strong gravitational fields strongly influence WPs decoherence. 

Ref.\cite{Petruzziello:2020wea}  has considered the implications of these results for an inertial observer. Further studies should address this question, by including the coupling with matter and neutrinos, outside the compact object. 

\begin{figure}\label{fig:curved}
\begin{center}
\includegraphics[width=.35\textwidth]{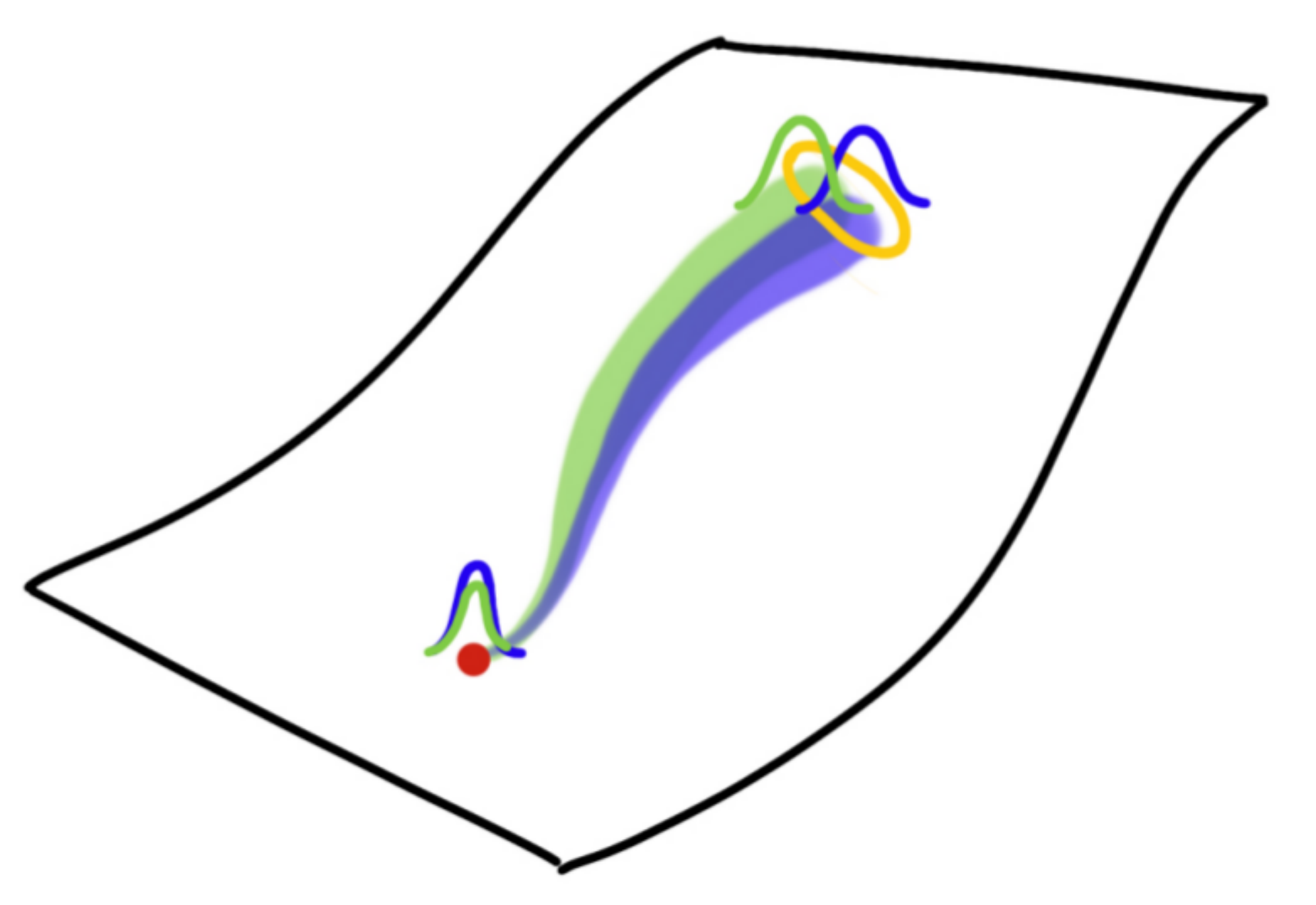}
\includegraphics[width=.3\textwidth]{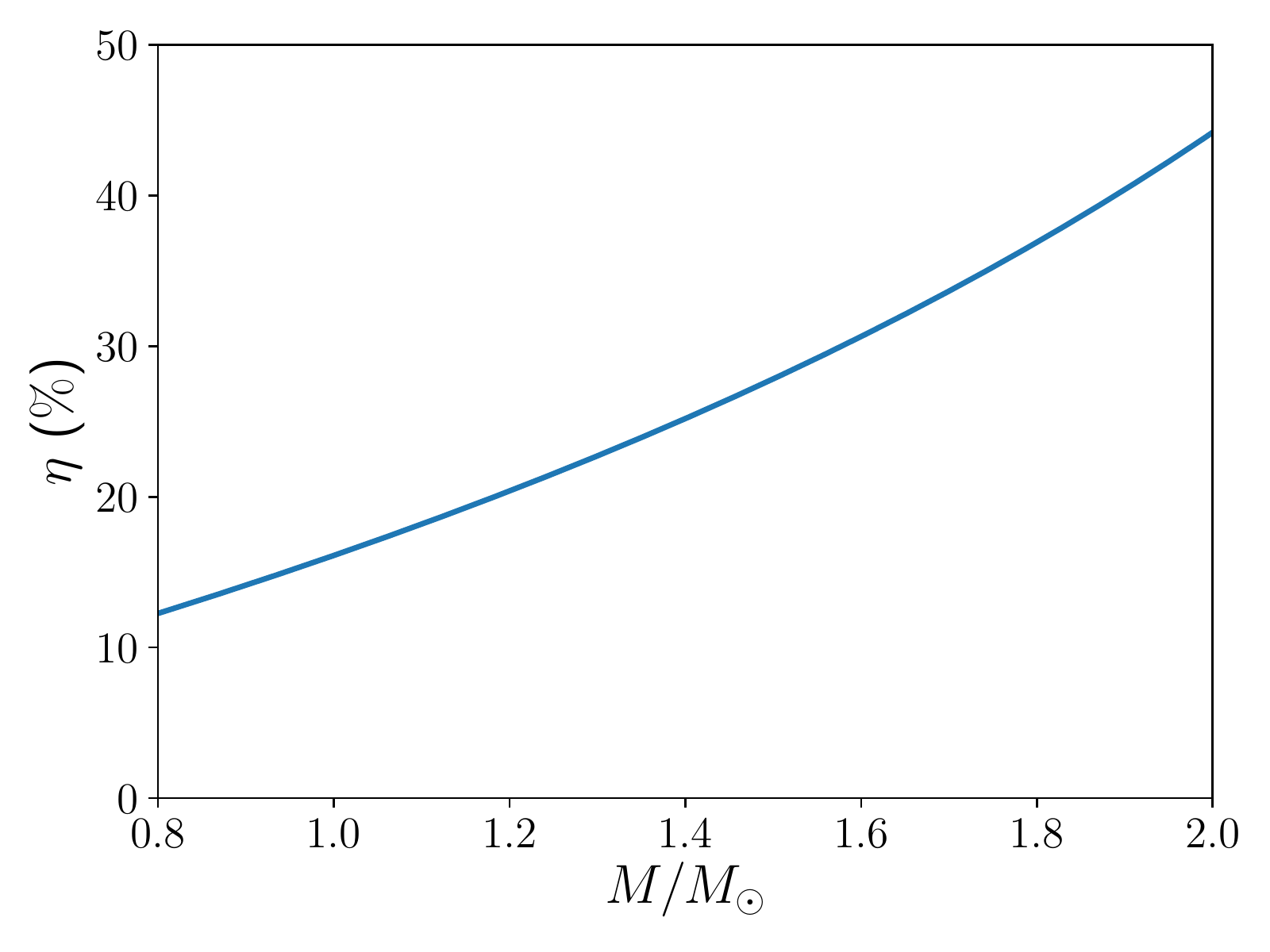}
\caption{Left : Artistic drawing of neutrino WPs propagating from a production point P to "detection" point D in presence of strong gravitational fields. Each mass eigenstate WPs follows a trajectory close to null-geodesics. The coloured widths picture the distribution of trajectories due to the WP finite extension. Right : Relative difference between the coherence proper time (curved) and length (flat spacetime), as a function of the Schwarzschild mass $M$. Adapted from \cite{Chatelain:2019nkf}.}
\end{center}
\end{figure}

\section{Future observation of the diffuse supernova neutrino background}
Neutrinos from past supernova explosions constitute the relic, or diffuse, supernova neutrino background (DSNB). The Super-Kamiokande Collaboration has started taking data with the addition of Gadolinium that should  improve the signal-to-background ratio by improving neutron tagging \cite{Beacom:2003nk}. The JUNO experiment is currently under construction \cite{JUNO:2015zny}, while the Hyper-Kamiokande has been approved \cite{Abe:2011ts}. Available predictions are close to the current upper bounds \cite{Super-Kamiokande:2013ufi,Super-Kamiokande:2021acd} indicating the DSNB discovery should be close.  

The DSNB is sensitive to the core-collapse supernova rate which is still uncertain, the debated fraction of failed supernovae \cite{Lunardini:2009ya}, non-standard neutrino properties (see e.g. \cite{DeGouvea:2020ang}), the contribution from binary systems \cite{Kresse:2020nto,Horiuchi:2020jnc}
and to redshifted supernova neutrino fluxes.  The latter are not only influenced by the MSW effect (usually implemented in predictions), but also by other flavor mechanisms. 

Ref.  \cite{Galais:2009wi} has investigated the combined effect of shock waves and neutrino self-interactions (in the so called "bulb" model). The results have shown that the spectral swapping produced by $\nu\nu$ interactions is counterbalanced by the phase effects due to shock waves. This interplay is shown to impact the expected DSNB rates in water Cherenkov  and argon based detectors such as DUNE by as much as 10-20$\%$, depending on the (inverted or normal) mass ordering, which is as sizeable as the MSW effect.        

Clearly, the upcoming discovery of the DSNB will bring invaluable information on supernova neutrino fluxes, neutrino properties, the core-collapse supernova rate and the fraction of failed supernovae. And maybe, we will be lucky enough to detect supernova neutrinos from a(n) (extra)galactic supernova one day.

\end{document}